\documentclass[letter, 12pt]{article}
\usepackage{arxiv}

\usepackage[utf8]{inputenc} 
\usepackage[T1]{fontenc}    
\usepackage[colorlinks]{hyperref}
\hypersetup{
    colorlinks=true,
    urlcolor=blue,
    citecolor = blue,
    allcolors = blue}
\usepackage{hhline}
\usepackage{dblfloatfix}
\usepackage{url}            
\usepackage{booktabs}       
\usepackage{amsfonts}       
\usepackage{nicefrac}       
\usepackage{microtype}      
\usepackage{lipsum}		
\usepackage{graphicx}
\usepackage{natbib}
\usepackage{doi}
\usepackage{caption}
\usepackage{multirow}
\usepackage{makecell}
\usepackage{amsmath}
\usepackage{amssymb}
\usepackage{algorithmicx}
\usepackage{algorithm}
\usepackage[noend]{algpseudocode}
\usepackage{comment}
\usepackage{adjustbox}
\usepackage{tcolorbox}
\usepackage{bm}
\usepackage{float}
\usepackage{setspace}
\usepackage{longtable}

\usepackage{mathptmx}
\usepackage[T1]{fontenc}

\title{Personalized Student Attribute Inference}

\author{\href{https://orcid.org/0000-0001-9376-2642}{\includegraphics[scale=0.06]{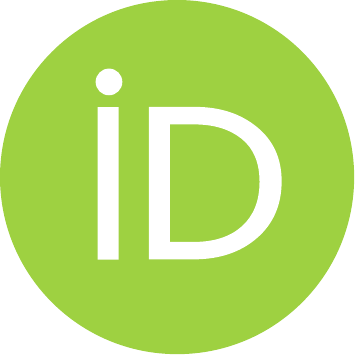}\hspace{1mm}Khalid Moustapha Askia}, 
\href{https://orcid.org/0000-0001-8196-2153}{\includegraphics[scale=0.06]{orcid.pdf}\hspace{1mm}Marie-Jean Meurs} \\
	Université du Québec à Montréal\\
	Montreal, QC, Canada\\
	\texttt{moustapha\_askia.khalid@courrier.uqam.ca}\\
	\texttt{meurs.marie-jean@uqam.ca} \\
}

\date{}

\renewcommand{\shorttitle}{Personalized Student Attribute Inference}

\hypersetup{
pdftitle={Personalized Student Attribute Inference},
pdfsubject={cs.AI},
pdfauthor={Khalid Moustapha Askia, Marie-Jean Meurs},
pdfkeywords={Big data, Educational data mining, Knowledge tracing, Machine learning},
}

\begin{document}
\maketitle

\begin{abstract}
Accurately predicting their future performance can ensure students successful graduation, and help them save both time and money.  
However, achieving such predictions faces two challenges, mainly due to the diversity of students' background and the necessity of continuously tracking their evolving progress. 
The goal of this work is to create a system able to automatically detect students in difficulty, for instance predicting if they are likely to fail a course.
We compare a naive approach widely used in the literature, which uses attributes available in the data set (like the grades), with a personalized approach we called \textit{Personalized Student Attribute Inference} (PSAI). 
With our model, we create personalized attributes to capture the specific background of each student.
Both approaches are compared using machine learning algorithms like decision trees, support vector machine or neural networks.
\end{abstract}


\section{Introduction}
\label{sec:intro}

As all academic institutions aim to improve the quality of education, the success of their students is essential. 
To make university affordable and worthwhile, it is hence important to ensure that most of the students enrolled in a program succeed it and graduate on time. 
Therefore, early interventions for students who most likely will fail their courses can help them save both time and money.
A possible solution towards this end is to build an automatic system that would successfully predict their future outcome.  
However, predicting students' performance is complex.  
The attributes frequently used by researchers are the Grade Point Average (GPA), internal assessment and students' demographic (gender, age etc). 
The issue with those attributes is that they tell nothing valuable about the student background and what s/he has been through. 
For that reason,  predicting methods  need  to  incorporate a way to capture students' background along with the historical student  accomplishments (grades, credits obtained, GPA).

We developed a personalized model called \textit{Personalized Student Attribute Inference} (PSAI), which creates personalized attributes to capture the specific background of each student. 
We compare our model with a naive approach, which uses directly the attributes available in the data set (like the grades, credits obtained, GPA, etc.). 
The next Section explains our approach.
Section~\ref{sec:exp} describes our experimental process and results, and finally, Section~\ref{sec:conclusion} discusses limitations and concludes.

\section{Personnalized Models}
\label{section:proposition}

We focus on personalized models, which take into account as much as possible the specifics of each student profile, and therefore emphasizes the student's background.
Grades, GPA, credits obtained, etc. are not sufficient since they cannot model a student's knowledge.
For example, students SA and SB have both a GPA of 3.7 but SA took only easy courses (3 courses in total) and SB took the most difficult ones (5 courses in total). 
Both have the same GPA so we cannot automatically determine who is the most talented. 
It shows that static attributes (the ones that are recorded directly like the grades) do not actually provide much profile details about a student. 
Thus, for accurately predicting student performance, one should consider other attributes as for instance the difficulty of the courses. 
To do so, we grouped courses by similar level of difficulty then we assigned them a weight, increasing with difficulty.
Also, we assigned a score to each student depending on the total number of courses s/he took, their difficulty and the grades s/he obtained. 

\subsection{Personalized Student Attribute Inference (PSAI)}

For assigning weights to courses according to their difficulty, we take a scale which limits are the weights of "extremely" easy courses and those of courses "extremely" difficult. 
We experimentally assign weights as follows: "extremely" easy courses get a weight of 0.5 and "extremely" difficult courses get a weight of 2.
An "extremely" difficult course is hence 4 times more difficult than an "extremely" easy course. 
By analyzing University marks system (where our data came from), for the "extremely" easy courses , we consider an average mark of 4.15 (between A (4.0) and A + (4.3) ) and for the "extremely" difficult courses, an average of 1.15 (between D (1.0) and D + (1.3)) .

Subsequently, in order to be able to assign a weight to a course according to the average mark obtained, we must consider a parametric function that will take this average mark as input and output the associated weight in accordance with the limits established above.
The function must also be decreasing, \textit{i.e.} if the input (the average mark) increases, the output (the weight) must necessarily decrease. 
Let $ \beta\times\exp{(-\alpha x )}$ be an exponential function where $\beta$ and $\alpha$ are parameters to be determined, and $x$ is the average of the marks obtained by the students who took the course. 
To estimate the parameters $\beta$ and $\alpha$, we use the limits we fixed. 

Solving the following equations system:
\[\left\{\begin{matrix} \beta\times\exp{(-1.15\alpha )} = 2 \\ \beta\times\exp{(-4.15\alpha )} = 0.5 \end{matrix}\right.\]
provides: 
$\alpha = \frac{\ln(4)}{3}$ and $\beta =2\exp{(\frac{1.15\ln(4)}{3})} $ \\

Making use of this function that assigns a weight to a course according to its difficulty, we present hereafter our algorithm to create a personalized data set, which will be used to train machine learning algorithms and make performance predictions.

\subsection{PSAI Algorithm}

\begin{algorithm}
\caption{PSAI Algorithm for course $A$}\label{euclid}
\begin{algorithmic}[1]
\Statex \textbf{Input:} Prior information on courses (average mark in the course, marks obtained) took by students that took course $A$
\State $\alpha \gets \frac{\ln(4)}{3} $
\State $\beta \gets 2\exp{(\frac{1.15\ln(4)}{3})}$
\State \textbf{For} each student that took course $A$: 
\State  \hspace{0.5cm} \textbf{For} each course $i$ taken before course $A$:
\State  \hspace{1cm} Let $ m_{i} $ be the average mark in the course $i$ (according to all students that took this course)
\State  \hspace{1cm} Let $ n_{i} $ be the mark of the current student in course $i$
\State  \hspace{1cm} Compute the score  of the student in the course i: $S_{i} = n_{i}\times\beta\times\exp{(-\alpha m_{i})} $
\State  \hspace{0.5cm} \textbf{End For}
\State  \hspace{0.5cm} Compute the total score of the student: $S$ = mean of $S_{i}$
\State  \textbf{End For}
\State Compute the weight of the course $A$: $ W_{A} = \beta\times\exp{(-\alpha m_{A})} $ where $m_{A}$ is the average mark in course $A$
\Statex \textbf{Output: } A score for each student and the computed weight for course $A$
\end{algorithmic}
\label{algo}
\end{algorithm}

Algorithm \ref{algo} provides a dataset that will be used to train the prediction model.
This dataset contains a score for each student and the weight of the course for which we want to predict the student performance. 
We also add as an attribute the overall success rate in the course. %

\section{Experiments and Results}
\label{sec:exp}

For the sake of brevity, we present our results only for the following course (acronyms changed for non-disclosure reasons): ABC2222 : 6483 students  with 5256  success and 1227 failures.
The question asked to our models is the following: \textit{Will a given student fail the course?}

The method of training and testing in  our experiments is the cross validation~\cite{kohavi1995study}. 
We used the following machine learning algorithms : Decision trees~\cite{safavian1991survey},  K-Nearest Neighbors~\cite{cover1967nearest}, Support Vector Machine (SVM)~\cite{cortes1995support}, Random Forest~\cite{breiman2001random}, an ensemble learning model (AdaBoost)~\cite{dietterich2002ensemble} and Neural Network~\cite{sarle1994neural}.
The evaluation metric is the F-measure (or F1-score)~\cite{sasaki2007truth}.

Table~\ref{tab:res} shows the results obtained with several machine learning algorithms. 
We compare our results with those obtained using a direct (naive) method (the standard method) which only uses the attributes related to the course and the students present in the database. 
In our case these attributes are: admission base, citizenship, previous program, legal status, college program,  age, gender, number of course credits obtained and GPA.

\begin{table}[H]

\centering
    \renewcommand{\arraystretch}{1.5}
\caption{Failure prediction in ABC2222 results}
\begin{tabular}{|l|l|l|}
\hline
\multicolumn{3}{|c|}{ABC2222}
\\ \hline
\multirow{2}{*}{Algorithm}  & Naive method & PSAI  \\ \cline{2-3} 
                            & F-measure(\%)  & F-measure(\%)  \\ \hline
Neural network          & 32,87            & 68,47            \\ \hline
Decision tree           & 44,37            & 66,37            \\ \hline
Adaboost                    & 45,70            & \textbf{69,57}            \\ \hline
k-NN                        & 37,07            & 64,31            \\ \hline
Random Forest               & \textbf{47,36}             & 58,30            \\ \hline
SVM                         & 27,81            & 62,75            \\ \hline
\end{tabular}
\label{tab:res}
\end{table}
      

\section{Conclusion}
\label{sec:conclusion}

PSAI outperforms standard methods.
Despite the imbalanced nature of the data set including fewer failures than successes, our approach detects many of these failures. 
It proves the need to consider several "hidden" aspects including the difficulty of the courses taken. %
The main limitation of the approach is related to the data set itself. 
Some essential data are missing to improve the model. %
We do not have the information of students before their first registration at the university.
Hence, our model can only be used from the second course taken by the student.
In addition, as often when dealing with real data collected over a long period of time, a large part of the records is unusable because of missing data and errors/noises.

\noindent \textbf{Reproducibility.~~} This project is publicly released as an open source software in the following repository:\\{\small{\url{https://gitlab.labikb.ca/khalid/psai}}}\\

\noindent \textbf{Acknowledgment.~~} We acknowledge the support of the Natural Sciences and Engineering Research Council of Canada (NSERC), [MJM Canada NSERC Grant
number 06487-2017]

\bibliographystyle{unsrtnat}
\bibliography{refcai2020educ}

\end{document}